\documentclass[twocolumn,aps,pra,superscriptaddress]{revtex4}
\usepackage{graphicx,amsmath,amssymb}
\usepackage{dcolumn,fancyhdr}
\usepackage{bm,natbib,mathrsfs}
\usepackage{times}
\usepackage{txfonts,palatino}
\usepackage{epstopdf}
\usepackage{latexsym}
\usepackage{epsfig}
\usepackage{bbold}
\usepackage[colorlinks, linkcolor=blue, citecolor=blue, urlcolor=blue, breaklinks=true]{hyperref}

\newcommand{\fref}[1]{Fig.~\ref{#1}}

\newcommand{\rmd}{\text{d}}
\DeclareMathOperator{\Tr}{Tr}
\newcommand{\ket}[1]{\lvert#1\rangle}

\makeatletter
\newlength \figurewidth
\makeatother

\begin{document}

\title{Macroscopicity in an optomechanical matter-wave interferometer}
\author{Andr\'e Xuereb}
\address{Centre for Theoretical Atomic, Molecular and Optical Physics, School of Mathematics and Physics, Queen's University Belfast, Belfast BT7\,1NN, United Kingdom}
\address{Department of Physics, University of Malta, Msida MSD\,2080, Malta}
\author{Hendrik Ulbricht}
\address{School of Physics and Astronomy, University of Southampton, Southampton SO17 1BJ, United Kingdom}
\author{Mauro Paternostro}
\email{m.paternostro@qub.ac.uk}
\address{Centre for Theoretical Atomic, Molecular and Optical Physics, School of Mathematics and Physics, Queen's University Belfast, Belfast BT7\,1NN, United Kingdom}
\date{\today}

\begin{abstract}
We analyse a proposal that we have recently put forward for an interface between matter-wave and optomechanical technologies from the perspective of \emph{macroscopic quantumness}. In particular, by making use of a measure of macroscopicity in quantum superpositions that is particularly well suited for continuous variables systems, we demonstrate the existence of working points for our interface at which a quantum mechanical superposition of genuinely mesoscopic states is achieved.  Our proposal thus holds the potential to affirm itself as a viable atom-to-mechanics transducer of  quantum coherences.
\end{abstract}

\maketitle

Optomechanical devices are frequently alluded to as paving the path towards achieving macroscopic superpositions. This point of view is clearly motivated by the fact that the devices invoked in optomechanics, most commonly in the form of mirrors in a harmonic potential, may be ``macroscopic'' in the conventional sense; i.e., visible without the use of high-powered microscopes and composed of a macroscopic number of particles. However, the large physical dimensions of the systems at hand does not necessarily guarantee, \emph{de facto}, macroscopic character of the state one of them is prepared into. This point was beautifully elucidated by Leggett in his 1980 work, with the introduction of the concept of ``disconnectivity'' as a semi-quantitative instrument to argue that phenomena commonly intended as evidences of quantum effects at the macroscopic scale (such as the Josephson effects in superconductors or non-classical rotational inertia in superfluids) involve, in reality, only one- or two-particle processes~\cite{Leggett1980}. 

The problem of defining rigorous quantitative tools to characterise the degree of macroscopicity of a given quantum state has since then attracted some interest, most noticeably epitomised by the formulation of criteria that either refer the state of a system to specific representations or count the resources that are actually necessary to study such states~\cite{Dur2002,Bjork2004,Cavalcanti2006,Marquardt2008b,Lee2011c}. In Ref.~\cite{Nimm2013}, a measure built with the goal to exclude minimally invasive modifications of quantum mechanics that produce classical behavior at the macroscale has been put forward and applied to some relevant experimental examples.

In this paper we adopt the measure proposed in Ref.~\cite{Lee2011c} to address quantitatively the macroscopic character of the state produced by a recently proposed interface between matter-wave resources and an optomechanical ``detection stage''~\cite{Xuereb2013b}. We determine the working point at which a certifiable macroscopic quantum superposition is prepared  on the mechanical subsystem using the quantum coherence brought about by the matter-wave resource. %we propose an original use of such measure of macroscopicity in the context of revealing gravity-induced decoherence. Adsorption of molecules or similar particles on surfaces scrambles the internal state of the particles; our system is therefore insensitive to the kind of effects put forward in Ref.~\cite{Zych2011}, whereby general-relativistic time-dilation is encoded in the internal state of the particle, thereby `orthogonalising' the wave-function and degrading visibility. 
However, any relative phase picked up by one arm of the interferometer will carry forward to the mechanical subsystem and can be picked up in the off-diagonal terms of the density matrix. Shot-to-shot variations of this phase will appear as an effective decoherence mechanism whose effect on the size of the superposition state we investigate in broad terms in Sec.~\ref{sec:Decoherence}.
% {\bf MAURO: discussion to be enriched later on}

The remainder of this paper is organised as follows. In Sec.~\ref{sec:Model} we illustrate the basic principles of the matter-wave-to-optomechanics interface that embodies the technological platform for our study. Sec.~\ref{sec:Quantifying} is devoted to a brief introduction of the measure of macroscopicity that represents the main tool for our analysis and which is put in place to quantitatively characterise the state of the mechanical modes in the optomechanical detection stage of our proposal. Sec.~\ref{sec:Controlling} presents a study of the effects that the key parameters entering the formal description of the interface have on the degree of macroscopicity of the mechanical state, while in Sec.~\ref{sec:Decoherence} we provide an %starting from the 
analysis of standard environmental decoherence. %, we propose an interesting link between the inference of gravity-induced dephasing and the diagnostic capabilities of our setup. 
Finally, in Sec.~\ref{sec:Conclusions} we draw our conclusions and discuss open questions relevant to the study here at hand. 

\begin{figure}[t]
 \centering
    \includegraphics[width=0.9\figurewidth]{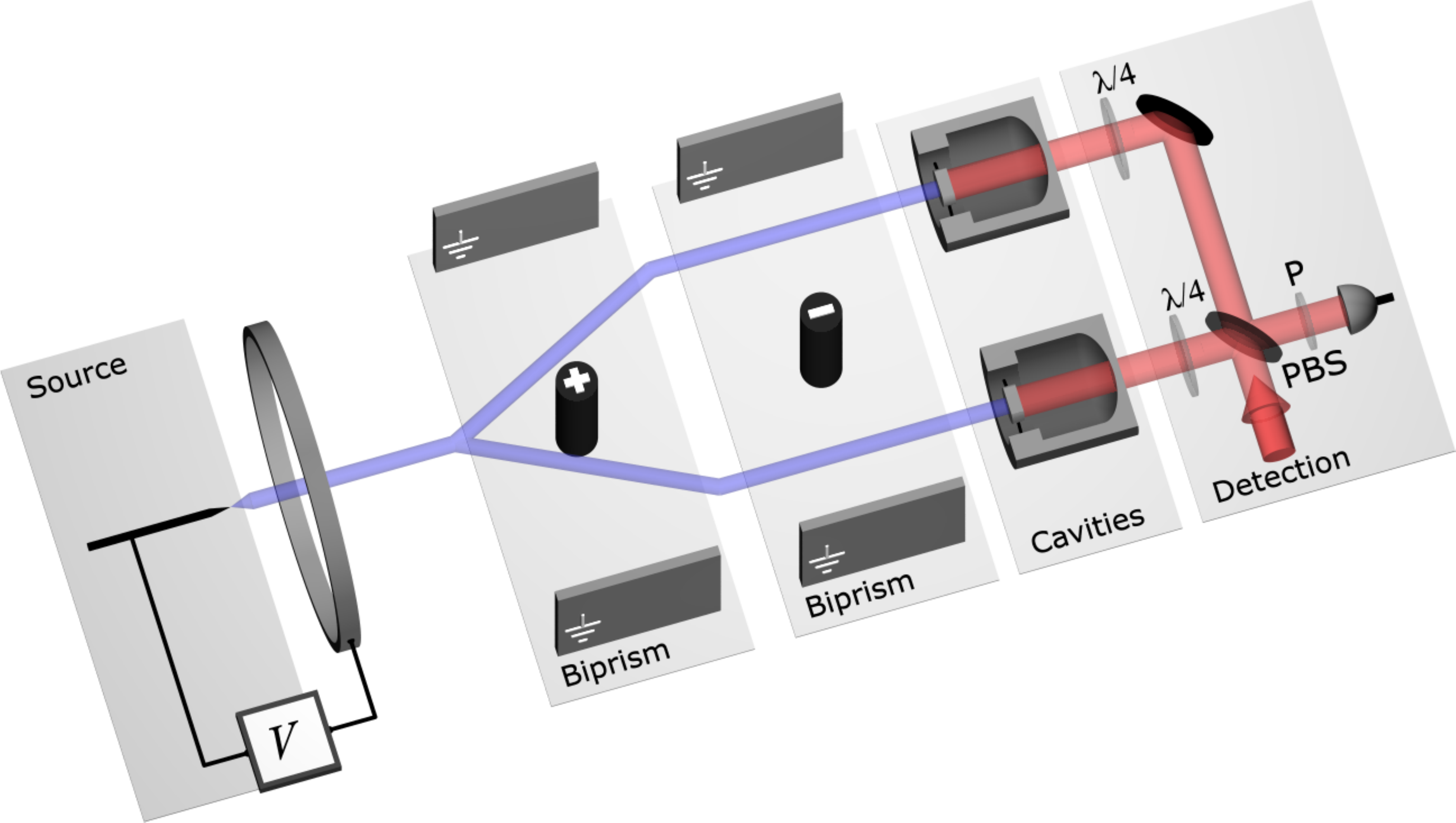}
 \caption{The experimental setup we explore. A matter wave is coherently split into two, and each arm of this matter-wave interferometer leads to an optomechanical cavity. Optical tomography is used to produce phase-space portraits of the joint mechanical state. Figure reproduced from Ref.~\cite{Xuereb2013b}.}
 \label{fig:System}
\end{figure}

\section{The model}
\label{sec:Model}

\noindent
The system we are interested in, as illustrated in \fref{fig:System}, is composed of two optomechanical cavities coupled to (i)~an optical interferometer used to drive the cavities and read out the state of the mirrors; (ii)~two arms of a matter-wave interferometer. The combination of these two systems is the key feature of our setup, enabling as it does the transfer of coherence from the matter waves to the optomechanical subsystem. After $N$ particles have passed through the matter-wave interferometer, thus producing the initial matter-wave state $\ket{\psi}_{\rm part}=\binom{2N}{N}^{-1/2}\sum^N_{r=0}\binom{N}{r}\ket{N-r,r}_{\rm part}$, the state of the two mechanical mirrors can be described by the density matrix 
\begin{multline}
\rho_\mathrm{mech}=\frac{1}{\pi^2}\int \rmd^2\alpha^{(1)}\rmd^2\alpha^{(2)}d^2\mu^{(1)}d^2\mu^{(2)}\\
\times W(\alpha^{(1)},\alpha^{(2)})\Bigl[\otimes^2_{j=1}\hat O^{(j)}_{\mu^{(j)}}\Bigr]\,,
\end{multline}
with the phase-space variables $\alpha^{(j)}=\alpha^{(j)}_{\rm r}+\imath\alpha^{(j)}_{\rm i}$, $\mu^{(j)}\in\mathbb{C}$, the operator $\hat O^{(j)}_{\mu^{(j)}}=e^{\alpha^{(j)}\mu^{(j)*}-{\rm c.c.}}\hat D^{(j)}(-\mu^{(j)})$ proportional to the Weyl displacement operator $\hat D^{(j)}(\mu)$~\cite{Walls1995}, and the Wigner function~\cite{Xuereb2013b}
\begin{multline}
W(\alpha^{(1)},\alpha^{(2)})=\frac{4}{\pi^2(2\bar{n}+1)^2\binom{2N}{N}}\\
\times\sum_{r,r^\prime=0}^N\binom{N}{r}\binom{N}{r^\prime}\exp\biggl(-\frac{2}{2\bar{n}+1}\Bigr\{{\alpha^{(1)}_\mathrm{r}}^2+{\alpha^{(2)}_\mathrm{r}}^2\\
+\bigl[\alpha^{(1)}_\mathrm{i}-\tfrac{\gamma}{2}(2N-r-r^\prime)\bigr]^2+\bigl[\alpha^{(2)}_\mathrm{i}-\tfrac{\gamma}{2}(r+r^\prime)\bigr]^2\Bigr\}\biggr)\\
\times\cos\bigl[2\gamma(r-r^\prime)(\alpha^{(1)}_\mathrm{r}-\alpha^{(2)}_\mathrm{r})\bigr],
\end{multline}
The parameter $\gamma$ quantifies the momentum gained by one of the mirrors upon being hit by a particle, and $\bar{n}=\left\{\exp[\hbar\omega_\mathrm{m}/(k_\mathrm{B}T)]-1\right\}^{-1}$ is the average number of phonons in the thermal baths (temperature $T$ and Boltzmann constant $k_\mathrm{B}$) connected to the two mirrors, which are assumed to be identical and having oscillation frequency $\omega_\mathrm{m}$. Although it is in principle possible to have the mechanical mirrors being connected to baths at different temperatures, without affecting the generality or validity of our analysis, here we assume $T$ to be the same on the two arms of the interferometer.
\par
The question we want to ask in this paper is, therefore, \emph{How ``macroscopic'' is the superposition state formed in this system?} After addressing how this question may be answered, we shall briefly discuss how the size of this state depends on the various parameters that enter the model. Upon introducing decoherence, we uncover a nontrivial interplay between $\gamma$, which acts to make the superposition more macroscopic, and $\bar{n}$, which acts in the opposite way.

\begin{figure}[t]
 \centering
    \includegraphics[width=\figurewidth]{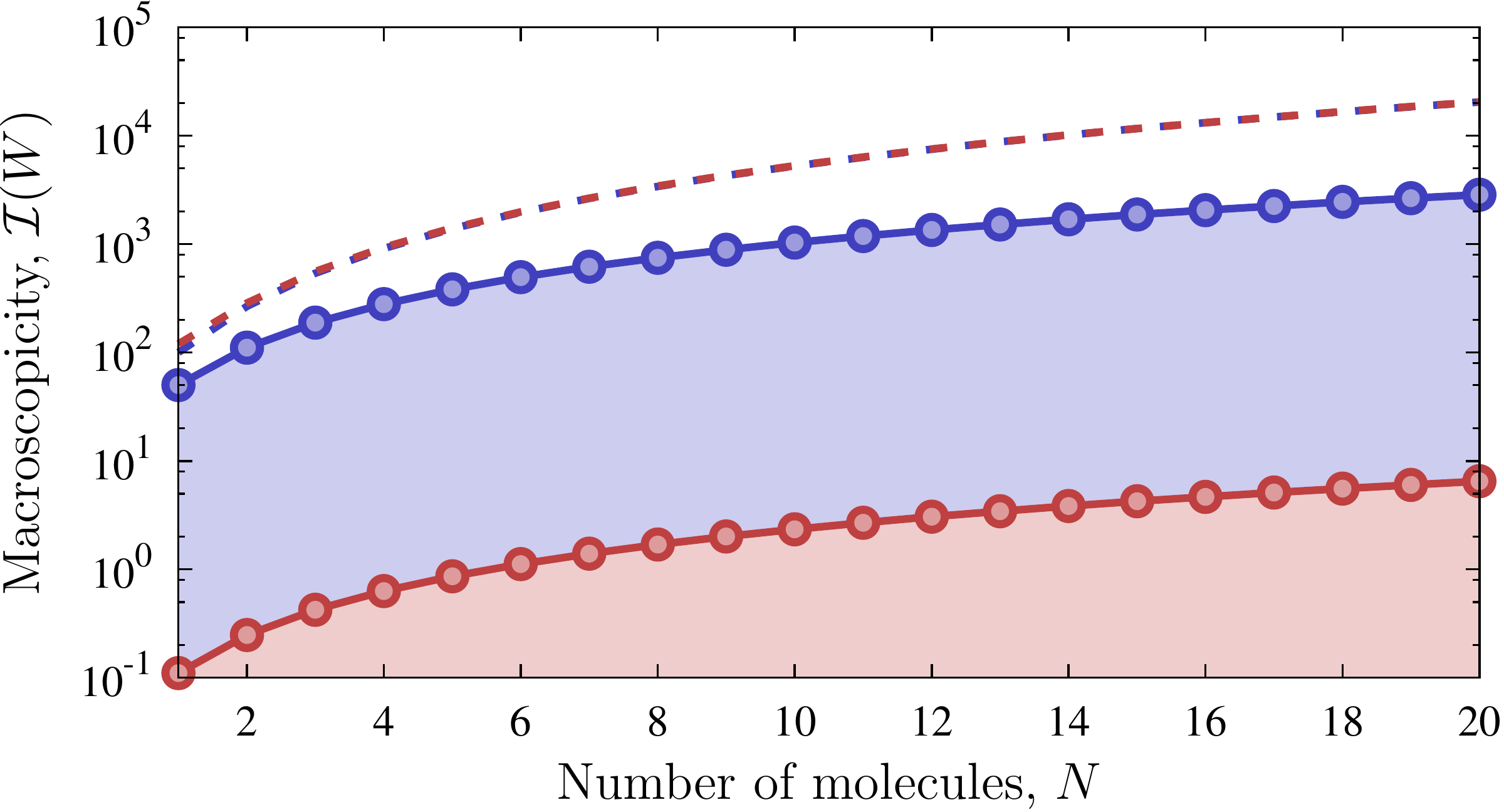}
 \caption{Effect of increasing the number of particles impinging on the mirrors for a constant value of $\bar{n}$; the macroscopicity of the superposition state increases monotonically with increasing $N$. We show two values for $\bar{n}$: $0$ (blue data points), and $10$ (red). For this figure we took $\gamma=10$. The dashed blue curve represents the mean number of phonons in the system, $n_\mathrm{ph}$; note that $\mathcal{I}(W)>n_\mathrm{ph}$ throughout.}
 \label{fig:Nn}
\end{figure}

\section{Quantifying macroscopicity}
\label{sec:Quantifying}

\noindent
We shall take as our measure for macroscopicity the inter\-ference-based measure defined in Ref.~\cite{Lee2011c}. In our notation, we write
\begin{multline}
\mathcal{I}(W):=\max\left[0,-\frac{\pi^2}{2}\int\rmd^2\alpha^{(1)}\rmd^2\alpha^{(2)}\,W(\alpha^{(1)},\alpha^{(2)})\right.\\
\left.\times\sum_{m=1}^2\biggl(\frac{\partial^2}{\partial\alpha^{(m)}\,\partial{\alpha^{(m)}}^\ast}+1\biggr)W(\alpha^{(1)},\alpha^{(2)})\right]\,.
\end{multline}
As shown in Ref.~\cite{Lee2011c}, in the absence of decoherence effects, $\mathcal{I}(W)$ for a single-mode bosonic field is equal to the mean number of excitations in the field in the case of a squeezed vacuum or the Schroedinger cat state $\ket{\psi}\varpropto\ket{\alpha}+\ket{-\alpha}$ where the kets refer to coherent states. Furthermore it can be shown that $\mathcal{I}(W)$ is upper-bounded by the number of excitations in the system. Inspired by these results, we shall compare $\mathcal{I}(W)$ to the average number of phonons created in the mechanical subsystem in our scheme. The latter is calculated as
\begin{equation}
n_\mathrm{ph}=\Tr\left[\left(\hat{a}^\dagger_1\hat{a}^{\vphantom{\dagger}}_1+\hat{a}^\dagger_2\hat{a}^{\vphantom{\dagger}}_2\right)\rho_\mathrm{mech}\right],
\end{equation}
where $\hat{a}_{m}$ the annihilation operator for phonons in mechanical subsystem $m=1,2$. Our data shows that, for $N=1$ and in the absence of thermal or decoherence effects, $\mathcal{I}\gtrsim n_\mathrm{ph}$, with the closest approach occurring when $N=1$ and $\bar{n}=0$. For large $\gamma$, $N=1$, and $\bar{n}=0$, we obtain $\mathcal{I}(W)\approx2n_\mathrm{ph}$, the approximation improving super-exponentially with increasing $\gamma$.

\begin{figure}[t]
 \centering
    \includegraphics[width=\figurewidth]{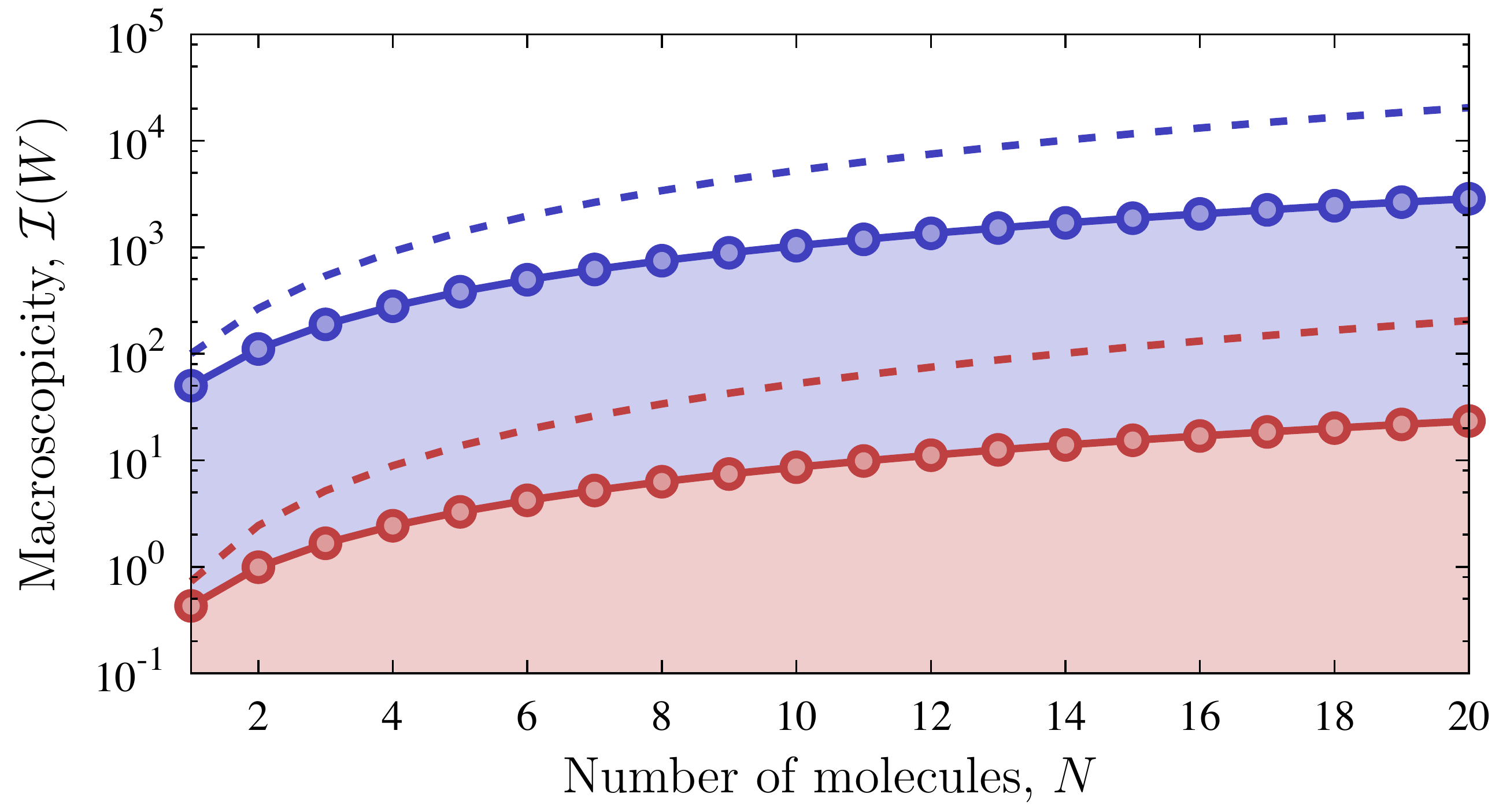}
 \caption{Effect of increasing the number of particles impinging on the mirrors for zero bath temperature. In this plot we show data for $\gamma=1$ (blue data points) and $\gamma=10$ (red). The dashed curves represent the mean number of phonons for the two respective situations.}
 \label{fig:Ngamma}
\end{figure}

Let us now specialise to our case. By making use of the definitions of $\alpha^{(m)}_\mathrm{r,i}$ ($m=1,2$), we can rewrite
\begin{multline}
\frac{\partial^2}{\partial\alpha^{(m)}\,\partial{\alpha^{(m)}}^\ast}=\frac{1}{4}\biggl[\frac{\partial^2}{\partial\alpha^{(m)}_\mathrm{r}\,\partial\alpha^{(m)}_\mathrm{r}}+\frac{\partial^2}{\partial\alpha^{(m)}_\mathrm{i}\,\partial\alpha^{(m)}_\mathrm{i}}\\
+\imath\biggl(\frac{\partial^2}{\partial\alpha^{(m)}_\mathrm{r}\,\partial\alpha^{(m)}_\mathrm{i}}-\frac{\partial^2}{\partial\alpha^{(m)}_\mathrm{i}\,\partial\alpha^{(m)}_\mathrm{r}}\biggr)\biggr]\,.
\end{multline}
We are now in a position to cast an explicit expression for $\mathcal{I}(W)$ into the form
\begin{equation}
\mathcal{I}(W)=\max\Biggl\{0,\frac{\sum_{r,r^\prime,R,R^\prime=0}^N\mathcal{N}(r,r^\prime,R,R^\prime)}{8(2\bar{n}+1)^2\bigl[\sum_{r,r^\prime=0}^N\mathcal{D}(r,r^\prime)\bigr]^2}\Biggr\}\,,
\end{equation}
with
\begin{equation}
\mathcal{D}(r,r^\prime):=\binom{N}{r}\binom{N}{r^\prime}e^{-(2\bar{n}+1)\gamma^2(r-r^\prime)^2}\,,
\end{equation}
and
\begin{multline}
\mathcal{N}(r,r^\prime,R,R^\prime):=\binom{N}{r}\binom{N}{r^\prime}\binom{N}{R}\binom{N}{R^\prime}\\
\times\exp\Bigl[-\tfrac{1}{2}(2\bar{n}+1)\gamma^2(r-r^\prime+R-R^\prime)^2\Bigr]\\
\times\Biggl\{\biggl[-\frac{8\bar{n}}{2\bar{n}+1}+\gamma^2\bigl(r-r^\prime+R-R^\prime\bigr)^2\biggr]\\
\times\exp\Bigl[2(2\bar{n}+1)\gamma^2(r-r^\prime)(R-R^\prime)\Bigr]\\
-\frac{8\bar{n}}{2\bar{n}+1}+\gamma^2\bigl(r-r^\prime-R+R^\prime\bigr)^2\Biggr\}\,.
\end{multline}
As is easily verified, when $\gamma=0$ we obtain $\mathcal{I}(W)=0$, as expected. We shall now proceed to explore the properties of $\mathcal{I}(W)$ as a function of the various parameters entering our model.

\section{Controlling macroscopicity}
\label{sec:Controlling}

\noindent
The setup we envisioned in Ref.~\cite{Xuereb2013b}, as explored above, has three parameters that determine fully the mechanical state formed after the interaction. Two of these parameters, $N$ and $\gamma$, work together to produce larger superposition states. By increasing $N$, the state formed will have an ever-increasing number of components. Similarly, if each particle has a larger effect on the motion of the mirrors (i.e., if we increase $\gamma$), the components of the superposition will be spaced further apart. In either case, we expect $\mathcal{I}(W)$ to increase monotonically with both $N$ and $\gamma$. On the contrary, $\bar{n}$ acts to spread out each peak in the superposition, gradually destroying the interference effects that distinguish a superposition state from a classical mixture. In this case, we expect $\mathcal{I}(W)$ to decrease monotonically with increasing $\bar{n}$. We shall discuss these two influences separately in the rest of this section.

\begin{figure}[t]
 \centering
    \includegraphics[width=\figurewidth]{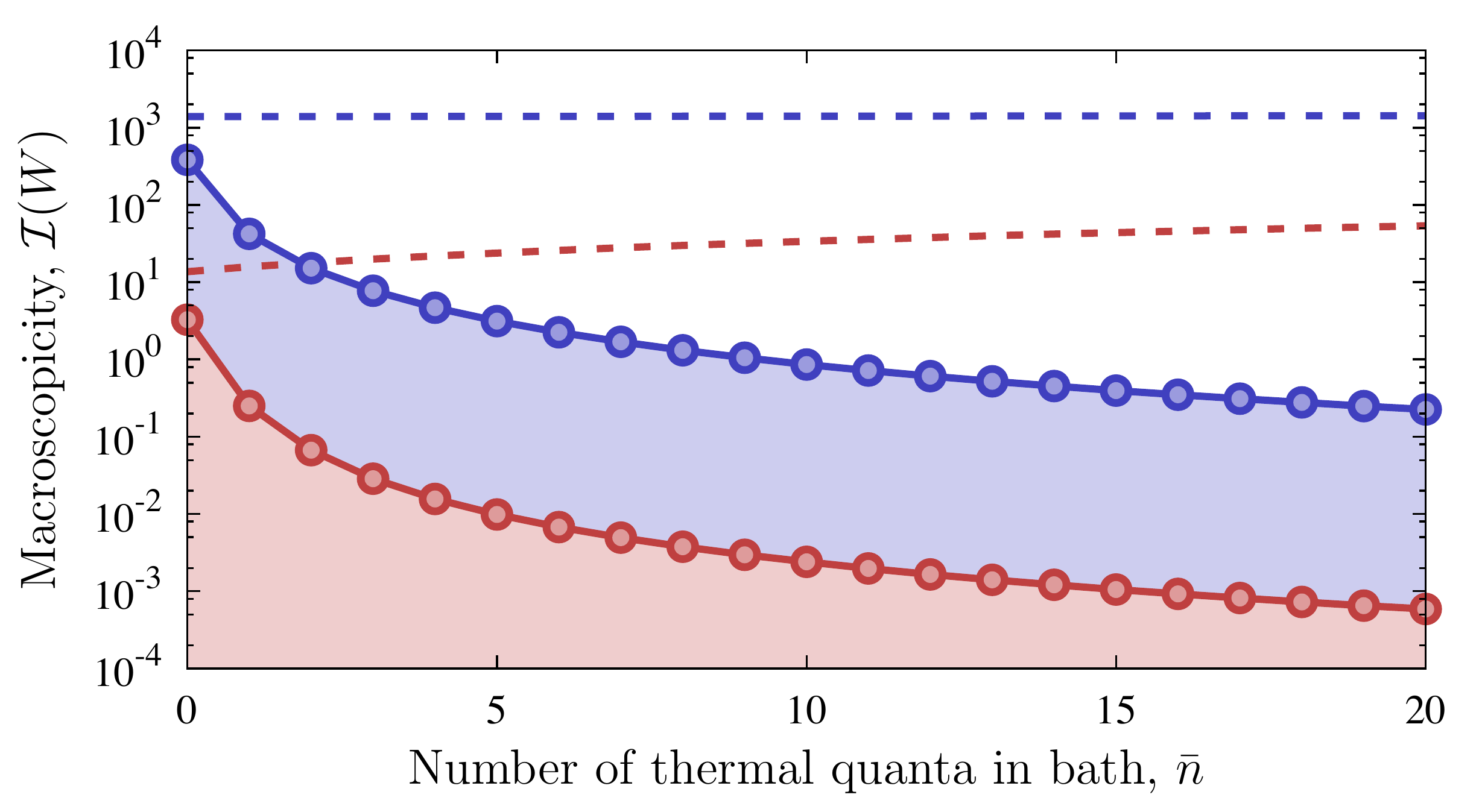}
 \caption{The opposing effects of $\gamma$ and $\bar{n}$. We show the situation for $N=5$ and plot $\mathcal{I}(W)$ as a function of the temperature of the mechanical baths. Increasing temperature leads to a decreased value for the size of the superposition state, but this may be counteracted by using a larger $\gamma$. The dashed curves represent the mean number of phonons. (Blue data points: $\gamma=10$, red: $\gamma=1$.)}
 \label{fig:ngamma}
\end{figure}
\subsection{Number of particles and interaction strength}
Our expectation is borne out entirely, with $\mathcal{I}(W)$ increasing dramatically as the number of particles is increased from $N=1$, in which case the two-mirror state is a simple two-component Schroedinger cat state shared between the two mirrors. This increase slows down somewhat as $N$ increases further, but no saturation is observed. We show this behaviour in \fref{fig:Nn}, where we plot the situation for two values of $\bar{n}$; we note that the two curves are displaced by approximately a constant factor. In this case we can see that states created with more components in the superposition (i.e., larger $N$) are more resilient to the detrimental effects of temperature. We note at this juncture that the upper bound of $\mathcal{I}(W)$ in the form of $n_\mathrm{ph}$ grows with $\bar{n}$, as expected, and therefore behaves in the opposite way to $\mathcal{I}(W)$ as the temperature is increased.
\par
Similarly, and as expected, the effect of increasing $\gamma$ is opposite to that of increasing $\bar{n}$. In \fref{fig:Ngamma} we show a dual plot to \fref{fig:Nn}: We fix the number of quanta in the baths to zero and plot data for two values of $\gamma$, demonstrating clearly that larger values of $\gamma$ create larger superposition states for any single value of $N$.

\subsection{Bath temperature}
The third parameter we investigate is the only one that is detrimental to the size of the superposition formed, as is clear from the data shown above. We demonstrate in \fref{fig:ngamma} that the effects of $\gamma$ and $\bar{n}$ run somewhat counter to one another: One may overcome the destruction of the superposition state brought about due to an increased temperature by using particles with a larger momentum (i.e., by increasing $\gamma$).

\begin{figure}[t]
 \centering
    \includegraphics[width=\figurewidth]{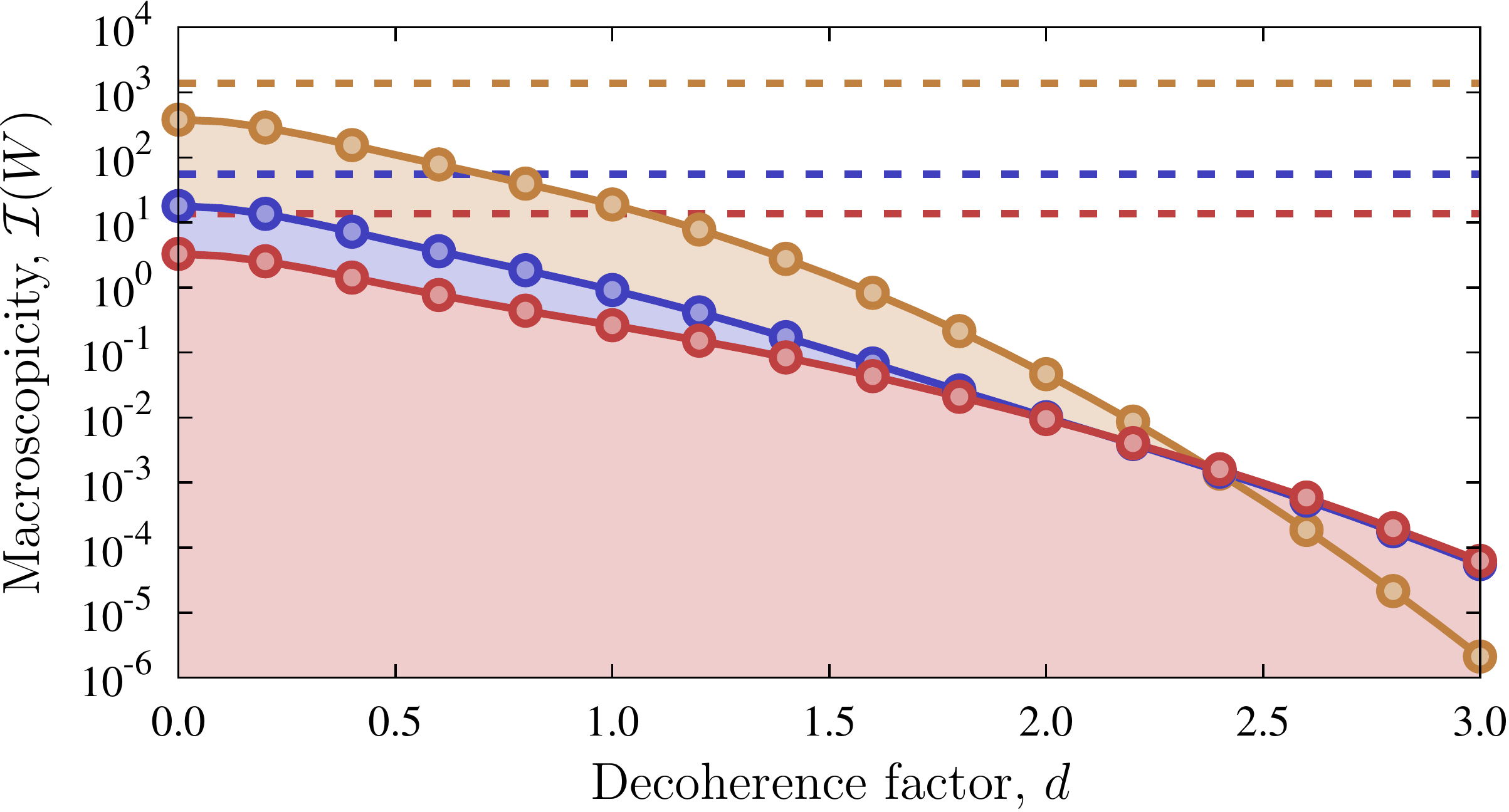}
 \caption{The effects of decoherence. We set $N=5$ and $\bar{n}=0$, and plot data for three values of $\gamma$: $1$ (red data points), $2$ (blue), and $10$ (brown). Note the cross-over point beyond which increasing $\gamma$ worsens the situation. The dashed curves represent the mean number of phonons at $d=0$.}
 \label{fig:d}
\end{figure}

\section{Decoherence}
\label{sec:Decoherence}

\noindent
We can model decoherence, without reference to a specific model, by introducing the function $\phi(r)=\exp\bigl[-(d\,r)^2\bigr]$, where we call $d$ the decoherence factor, which acts to destroy the coherences necessary to maintain a superposition state. To do this we replace
\begin{equation}
\mathcal{N}(r,r^\prime,R,R^\prime)\to\mathcal{N}(r,r^\prime,R,R^\prime)\phi(r-r^\prime)\phi(R-R^\prime)\,,
\end{equation}
and
\begin{equation}
\mathcal{D}(r,r^\prime)\to\mathcal{D}(r,r^\prime)\phi(r-r^\prime)\,,
\end{equation}
in $\mathcal{I}(W)$. By looking at the form of $W(\alpha^{(1)},\alpha^{(2)})$, one sees that the terms with $r\neq r^\prime$ originate from the off-diagonal terms in the two-mirror density matrix; it is precisely these terms to which the coherences between the matter waves in the two arms of the interferometer are transferred upon collision. We are thus describing effective dephasing-like noise, which we deem to be the most relevant for the sort of questions that we address in this work. Setting $d=0$ is equivalent to the decoherence-free model discussed thus far. One expects that increasing $d$ will act to reduce $\mathcal{I(W)}$, and indeed this is what we observe in \fref{fig:d}. However, we notice here a qualitative difference from the data set presented in \fref{fig:ngamma}; increasing $\gamma$ acts to counteract decoherence only for small $d$. For each value of $N$ (we show here only one) there exists some value of $d$ beyond which increasing $\gamma$ \emph{aids} decoherence. Whilst being insensitive to the mechanism producing the decoherence, in principle this fact allows us to use this setup to investigate decoherence and distinguish its effects from that of thermal excitations in the mechanical baths. Unfortunately, this mechanism is very delicate and quickly becomes overwhelmed by thermal effects.

\section{Conclusions}
\label{sec:Conclusions}

\noindent
We have addressed the macroscopic nature of the quantum state of a fully mechanical system prepared using a quantum interface able to transduce matter-wave coherence into mechanical quantum superpositions. We have analysed the behaviour of a recently introduced measure of macroscopicity of a quantum superposition designed to face the infinite dimensional nature of the Hilbert space of continuous variable systems. When compared to the average number of phononic excitations within the state of the mechanical system, such measure reveals that stringent experimental conditions should be met in order to certify unambiguously the macroscopic nature of the state engineered through the proposed interface. While providing important information for the experimental community interested in the tantalising challenge of extending the quantum framework to the mesoscopic and macroscopic domain, our results open up a series of questions. First, our analysis reveals the paramount importance of designing viable schemes for the engineering of macroscopic quantum states that are robust against temperature or decoherence effects. Second, our results may have unforeseen implications for the metrological estimation of gravitational decoherence effects: The strong dependence of the measure of macroscopicity on the mass of the mechanical systems, through $\gamma$, could make it a sensitive detector of gravity-induced decoherence, a possibility that we are currently exploring. 

\section*{Acknowledgements}
A.X.\ would like to thank the Royal Commission for the Exhibition of 1851 for financial support. H.U.\ acknowledges funding from the J.\ Templeton Foundation through the Foundational Questions Institute (FQXi), and the South-English Physics network (SEPNet). M.P.\ thanks the UK EPSRC for a Career Acceleration Fellowship and a grant awarded under the ``New Directions for Research Leaders'' initiative (EP/G004579/1), the John Templeton Foundation (grant 43467), and the EU Collaborative Project TherMiQ (grant egreement 618074).


\begin{thebibliography}{99}

\bibitem{Leggett1980} {Leggett, A. J.}, {Progress of Theoretical Physics Supplement} {\bf 69}, 80 (1980).

\bibitem{Dur2002}  {D\"{u}r, W., Simon, Ch., and Cirac, J.}, {Phys. Rev. Lett.} {\bf 89}, {210402} (2002).

\bibitem{Bjork2004} {Bj\"{o}rk, G. and Mana, P. G. L.}, {J. Opt. B} {\bf 6}, {429} (2004).

\bibitem{Cavalcanti2006} {Cavalcanti, E. and Reid, M.}, {Phys. Rev. Lett.} {\bf 97}, {170405} (2006).

\bibitem{Marquardt2008b} {Marquardt, F., Abel, B. and von Delft, J.}, {Phys. Rev. A} {\bf 78}, {012109} (2008).

\bibitem{Lee2011c} {Lee, C.-W. and Jeong, H.}, {Phys. Rev. Lett.} {\bf 106}, {220401} (2011).

\bibitem{Nimm2013} Nimmrichter, S. and Hornberger K., Phys. Rev. Lett. {\bf 110}, 160403 (2013).

\bibitem{Xuereb2013b} {Xuereb, A., Ulbricht, H., and Paternostro, M.}, {Sci. Rep.} {\bf 3}, {1} (2013).

\bibitem{Walls1995} {Walls, D. F. and Milburn, G. J.}, {\it Quantum Optics} (Springer, Heidelberg, 1995).

\end{thebibliography}
\end{document}